\documentclass[journal]{IEEEtran}
\usepackage{hyperref}
\usepackage{amsmath,amsfonts}
\usepackage{color}
\usepackage{bm}
\usepackage{threeparttable}
\usepackage{booktabs}
\usepackage{algorithmic}
\usepackage{cite}
\usepackage{array}
\usepackage{textcomp}
\usepackage{stfloats}
\usepackage{url}
\usepackage{verbatim}
\usepackage{eqnarray}
\usepackage{graphicx}
\usepackage{balance}
\usepackage{graphics}
\usepackage[usenames,dvipsnames]{xcolor}
\usepackage{dcolumn}
\usepackage[latin1]{inputenc}
\usepackage{textcomp}
\usepackage{footnote}
\usepackage{tablefootnote}
\usepackage{hologo}
\usepackage{booktabs}
\usepackage{epstopdf}
\usepackage{url}
\usepackage{amssymb}
\usepackage{csquotes}
\usepackage[british]{babel}
\usepackage{siunitx} 
\def\BibTeX{{\rm B\kern-.05em{\sc i\kern-.025em b}\kern-.08em
    T\kern-.1667em\lower.7ex\hbox{E}\kern-.125emX}}
\def \MgB2 {MgB$_{2}$ }

\begin{document}

\title{Practical Forecasting of AC Losses in Multi-layer 2G-HTS Cold Dielectric Conductors }
\author{M.~Clegg, and H. S. Ruiz\\College of Science and Engineering \& Space Park Leicester, University of Leicester, Leicester LE1 7RH, United Kingdom
\thanks{Manuscript submitted \today.}
\thanks{The authors are with the College of Science and Engineering $\&$ Space Park Leicester, University of Leicester, Leicester LE1 7RH, United Kingdom
(e-mail: \href{mailto:mlc42@leicester.ac.uk}{MC: mlc42@leicester.ac.uk}, \href{mailto:dr.harold.ruiz@leicester.ac.uk}{HSR: dr.harold.ruiz@leicester.ac.uk})}
\thanks{This work was supported by the UK Research and Innovation, Engineering and Physical Sciences Research Council (EPSRC), grant Ref. EP/S025707/1, led by H.S.R. All authors acknowledge the use of the High Performance Computing Cluster Facilities (ALICE) at the University of Leicester (UoL). M.C thanks the CSE and EPSRC-DTP studentships provided by UoL. Networking support provided by the European Cooperation in Science and Technology, COST Action CA19108 (Hi-SCALE), is also acknowledged.}
}

\markboth{XXXXXX,~Vol.~XX, No.~XX, October~2022}%
{Clegg \MakeLowercase{\textit{et al.}}: Practical Forecasting of AC Losses in Multi-layer 2G-HTS Cold Dielectric Conductors}

%
\maketitle


\begin{abstract}
With the recent progresses on the designing and manufacturing of lightweight and high engineering current density superconducting cables, the need for an established, fast, and sufficiently accurate computational model for the forecasting of AC-losses in cold-dielectric conductors, is pivotal for increasing the investment confidence of power grid operators. However, validating such models is not an easy task, this because on the one hand, there is a low availability of experimental results for large scale power cables and, on the other hand, there is a large number of 2G-HTS tapes involved whose cross-sectional aspect ratio hinders the numerical convergence of the models within reasonable delivery times. Thus, aiming to overcome this challenge, we present a detailed two-dimensional H-model capable to reproduce the experimentally measured AC-losses of multi-layer power cables made of tens of 2G-HTS tapes. Two cable designs with very high critical currents have been considered, the first rated at $1.7~\mathrm{kA}$ critical current, consisting of fifty $4~\mathrm{mm}$ width 2G-HTS tapes, these split in 5 concentric layers wound over a cylindrical former, with the three inner layers forming an arrangement of 24 tapes shielded by two further layers with 13 tapes each. This cable is contrasted with a size wise equivalent cable with67 superconducting tapes rated at $3.2~\mathrm{kA}$ critical current, whose design implies the use of 40 tapes of $3~\mathrm{mm}$ width split within four core layers, and 27 tapes of $4~\mathrm{mm}$ width distributed in two shielding layers. In both situations a remarkable resemblance between the simulations and experiments has been found, rendering to acceptable estimates of the AC-losses for cold dielectric conductors, and offering a unique view of the local electrodynamics of the wound tapes where the mechanisms of shielding, magnetization, and transport currents can coexist within the hysteretic process.
\end{abstract}




\IEEEpeerreviewmaketitle


\section{Introduction}\label{Sec.1}

\IEEEPARstart{A}{fter} the advent of the so-called second generation of high temperature superconducting tapes (2G-HTS), a greater interest on the development of power cables under direct (DC) and alternating (AC) current conditions has been seen~\cite{gilbertson2000electrical,PamidiS.2015HsHp,PrusseitWerner2015PASC,Doukas2019IEEE}. Although DC systems offer a further reduction in electrical losses, depending on the amount of current being transmitted~\cite{Ruiz2018aIEEE,Ruiz2018bIEEE,Ruiz2018SUST,Ruiz2012APL,Ruiz2013IEEE,Ruiz2013JAP}, most existing power grids are predominantly AC. Meaning any DC cable would require additional expenses in the conversion, which is not justifiable unless the cable is in excess of 100 miles \cite{Rey2015}. Likewise, compact HTS cables can be used within aircraft and ship propulsion systems but, both the mass and the dimensions of the cable need to be considered for their production~\cite{Fetisov2020}. One way that this can be achieved is by cable designs with multilayer structures of 2G-HTS tapes, which can simultaneously increase the transport current carrying capabilities alongside reducing the cross-sectional area of the cable.  In turn, resulting in the lightening of the cryogenic load \cite{Rey2015,Fetisov2020,Kalsi2011SC} and any associated costs through the materials \cite{Willen2005CIRED}.


\begin{center}
\begin{table}[htbp]
\centering\small
\caption{\label{Tab:1} Parameters$^\dag$ of the trilayer plus bilayer `shielded' single-phase cable in accordance with the prototype at~\cite{Fetisov2018IEEE}.}
\begin{tabular}{ccccccccc}
\toprule
$N_{l}$ & $W_{l}$ & $R_{l}$ [{mm}] & $\Gamma_{l}$ [{mm}] & $N_{T}$ & $\langle I_{c} \rangle$ [{A}]&\\
\midrule
&&&Cable Core&\\
\midrule
1st&4&11.3&+200&8&72.8&\\
2nd&4&12.2&+109&8&72.8&\\
3rd&4&13&+61&8&72.8&\\
\midrule
&&&Cable `Shield'&\\
\midrule
1st&4&18.4&-330&13&91.8&\\
2nd&4&19.6&+110&13&91.8&\\
\bottomrule
\end{tabular}
\begin{tablenotes}
       \item $^\dag$ \footnotesize{$N_{l}$ stands for the number of layers, $R_{l}$ for its inner diameter, and $N_{T}$ is the number of tapes (per layer). $\langle I_{c} \rangle$ corresponds to the averaged $I_{c}$ per tape per layer, where $W_{l}$ refers to the width of the tape, and $\Gamma_{l}$ is the twist pitch length with positive or negative winding direction.}
\end{tablenotes}
\end{table}
\end{center}

\begin{figure}[!t]
\centering
\resizebox{0.489\textwidth}{!}{\includegraphics{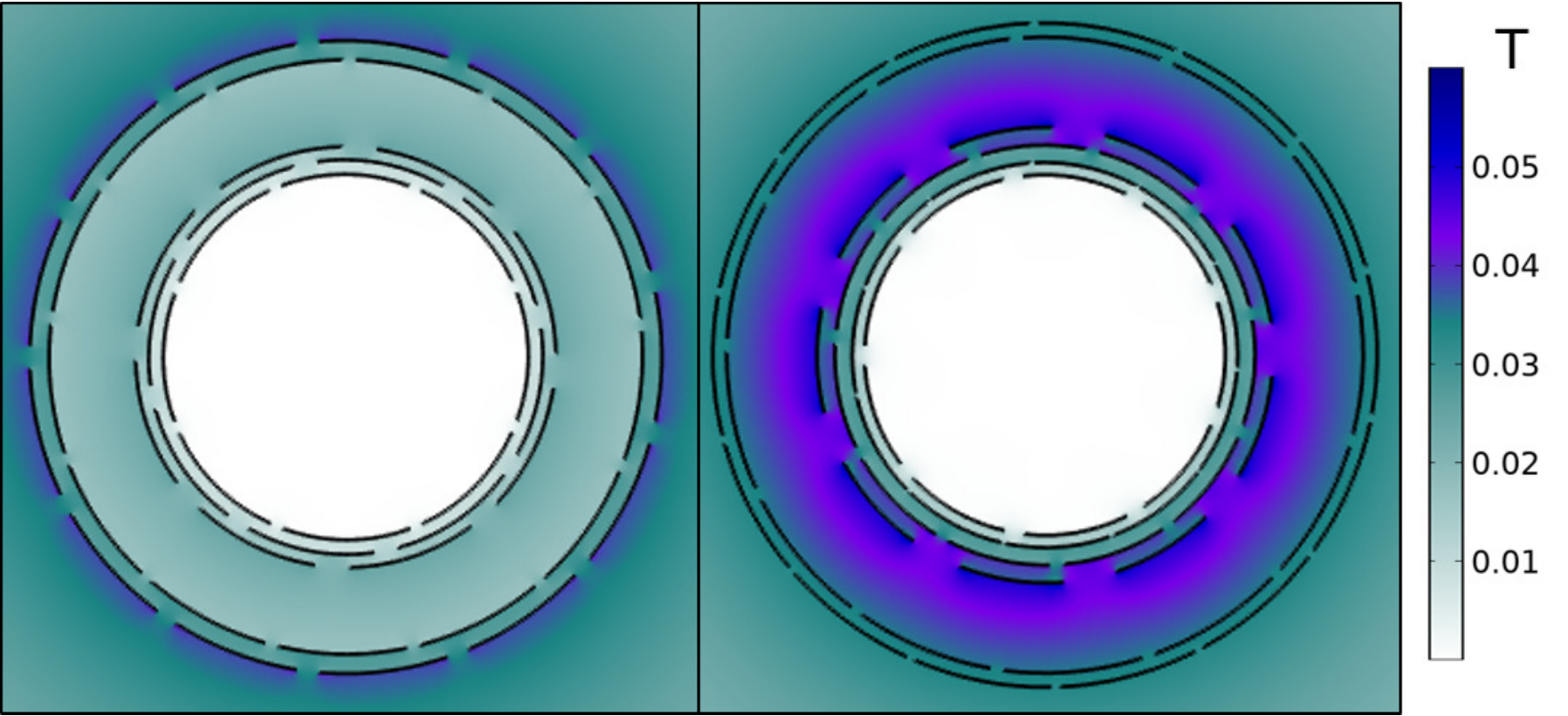}}
\caption{\label{Fig_1} {Two-dimensional representation of the prototype cables, \textit{(i)} at the left pane and \textit{(ii)} at the right, with the corresponding magnetic profiles at $t=0.025~\mathrm{s}$, i.e., when an applied alternating current $I_{tr}=0.5I_{c}\sin{(\omega t)}$ is at its peak value, being the frequency of the transport current $50~\mathrm{Hz}$.}
}
\end{figure}

With numerous HTS cables installed within research facilities and operational power grids \cite{PamidiS.2015HsHp,PrusseitWerner2015PASC,Doukas2019IEEE}, our recent work has taken a particular focus in the cables developed by SuperOx and the VNIIKP Cable institute, where a range of multilayer cable prototypes have been manufactured for both DC and single to three-phase AC systems \cite{Fetisov2018IEEE, Fetisov2020IOP, Fetisov2007IEEE, Fetisov2009IEEE, Fetisov2016IEEE, Fetisov2017IEEE, Fetisov2021IEEE}. For the case of AC cables, relevant research has been pursued within the context of cold dielectric designs, i.e., in cables where the outer layers of superconductors are ultimately aimed to serve as AC magnetic shields, looking to maintain the high current capacity of the cables, whilst making the prototypes to be more compact. Then, for benchmarking our numerical results we have chosen to reproduce and validate the experimental observations made in two of their most recent designs for single-phase power cables (see \autoref{Fig_1}), namely:

\begin{itemize}
    \item[\textit{(i)}] A total of 50 SuperOx tapes of $4~\mathrm{mm}$ width each are distributed in five concentric layers (\autoref{Fig_1} left-pane), following the design parameters shown in \autoref{Tab:1}. The manufacturing and initial experimental testing of this cable has been reported at~\cite{Fetisov2018IEEE}. Notice that, on the one hand, the cable `core' layers refer to the three inner layers with eight 2G-HTS tapes each, building up to $~\sim1.7~\mathrm{kA}$ critical current. On the other hand, the two outer layers consist of 26 tapes evenly distributed, which could be eventually used as a `shield' for the magnetic field created by neighboring single-phase cables, or add a further $~\sim2.3~\mathrm{kA}$ critical current if connected in series to the `core'. 
    \item[\textit{(ii)}] A size wise equivalent cable (\autoref{Fig_1} right-pane), whose design involves 67 SuperOx tapes distributed across the `core' and `shield` layers~\cite{Fetisov2020IOP}, but with the core rated critical current $(~\sim3.2~\mathrm{kA})$ nearly doubling the one for the cable design (i). In this case (see \autoref{Tab:2}), the core layers are made of 40, $3~\mathrm{mm}$ width, 2G-HTS tapes distributed across 4 concentric layers, which are surrounded by a further 2 layers (the shield) with 27,  $4~\mathrm{mm}$ width, tapes provided by the same manufacturer. 
\end{itemize}

On the one hand, regarding to the experimental conditions on which these cables have been tested~\cite{Fetisov2018IEEE,Fetisov2020IOP}, which therefore act as preconditions for our computational study, it is worth mentioning that for the cable \textit{(i)} the so-called `shield' layers are actually connected in series with the `core' layers. This means that the `shield' layers are configured to be under the same transport current conditions than the `core' layers, what blot out the intended purpose of magnetic shielding. However, the considered case allows to get a maximum estimation of the AC losses for these layers without the need for building a complex and costly rig with neighboring power cables, as the magnetization losses created by the called `shield' layers are not expected to be greater than the sum of their own transport current losses, plus the induced magnetization losses generated by the inner `core' layers~\cite{Ruiz2012APL,Ruiz2013JAP}. Simulations of the cable with the shield layers being disconnected have also been performed, i.e., without transport current, this in order to estimate the minimum AC losses for single or isolated power cables. 

On the other hand, regarding the reliability and accuracy of the chosen computational model, our group has extensively used the H-formulation in the past, already showing an excellent agreement between the numerical estimations and experimental measurements in a large set of power cables,  either in 2D or 3D environments  \cite{Ruiz2019MDPI,Ruiz2019JAP,Ruiz2019IOP,Ruiz2019IEEE,Clegg2022IOP}. Conclusions from these studies have shown that the simplified 2D models can be effectively used as digital twins of warm and cold dielectric conductors, allowing the timely assessment and forecasting of their power density losses. Still, we would like to emphasize that no matter how precise the 2D approach might result, which as a matter of fact depends on how realistic the physical parameters for individual tapes are (e.g., $J_{c}$ at the wound tapes and in the operational conditions of the entire cable), it is safe to say that a discrepancy with the experimental results of at least $\pm10\%$
is customarily considered as acceptable~\cite{PetrovJoP2020}.


\begin{center}
\begin{table}[!t]
\centering\small
\caption{\label{Tab:2} Parameters of the quad-layer plus bi-layer shielded single-phase cable$^{\ddag}$ in accordance with~\cite{Fetisov2020IOP}.}   
\begin{tabular}{ccccccc}
\toprule
$N_{l}$ & $W_{l}$ & $R_{l}$ [mm] & $\Gamma_{l}$ [mm] & $N_{T}$ & $\langle I_{c} \rangle$ [A]&\\
\midrule
&&&Cable Core&\\
\midrule
1st&3&10.32&-56.2&9&81\\
2nd&3&11.03&-193.6&11&81\\
3rd&3&12.03&94.3&11&81.4\\
4th&3&13.06&40.7&9&81.7\\
\midrule
&&&Cable Shield&\\
\midrule
1st&4&18.25&349.4&13&120\\
2nd&4&19.06&-317.4&14&118.6\\
\bottomrule
\end{tabular}
\begin{tablenotes}
       \item $^\ddag$ \footnotesize{See in conjunction with the cable \textit{(i)} definitions at \autoref{Tab:1}}
\end{tablenotes}
\end{table}
\end{center}


Thus, under the above considerations, in this paper we present reliable numerical estimations for large scale multilayer HTS power cables, based in our comparison with the experimental data of two of the most recent prototypes for cold dielectric cables produced by VNIIKP. Thus, our numerical approach is summarized in \autoref{Sec.2}, alongside with the disclosure of the electromagnetic profiles for individual 2G-HTS tapes, where general features for the local distribution of current density and profiles of magnetic field are informed. Further analysis of the AC losses is provided in \autoref{Sec.3}, with a depth analysis on the energy losses expected at each of the individual layers of the cable, and how our numerical predictions contrast well with the experimental observations. Finally, the main findings of our simulations will be briefed in \autoref{Sec.4}. 


\section{Numerical Implementation and Electromagnetic Modelling}\label{Sec.2}

Using the information in Tables \ref{Tab:1} and \ref{Tab:2}, the computer-aided design of the single-phased cables can be drawn in a 2D-approach with $3~\mathrm{mm}$ and $4~\mathrm{mm}$ width 2G-HTS SuperOx tapes~\cite{SuperOx}, with its magneto angular properties, $J_{c}(\mathbf{B},\theta)$ being characterized at \cite{Zhang2018}. The 1~\textit{$\mu$m} thickness of the superconducting layer is scaled by a factor of 50, improving thence the convergence and computing time of the model, as in previous works it has been already proven that a coherent rescaling of the self-field critical current density, allows for the proper reproduction of experimental results~\cite{Fareed2022IEEE,Ruiz2019MDPI}. Still, for an adequate computation of the local electrodynamics inside the superconducting layers, each of the scaled superconducting domains must be split in at least 7 sublayers, such that sufficient \textquote{computational resolution} is enabled for the convergence of Maxwell equations as shown in \cite{Clegg2022IOP}. Not such high resolution is required for any of the other components in the cable, as there is no current sharing nor eddy currents within them, i.e., behaving as electrical insulators which do not add any significant losses to the system. Still, it is necessary to incorporate its right dimensions for fidelity with the real system. 

About the computational formulation, the general form of the PDE module of COMSOL Multiphysics is used, where the state variable for the magnetic field $\mathbf{H}$ can be introduced by the partial derivatives function,
%
\begin{eqnarray}\label{Eq_1}
e_a\frac{\partial^2\bf{H}}{\partial t^2} + d_a\frac{\partial\bf{H}}{\partial t} + \nabla \cdot \Gamma= f \, . 
\end{eqnarray}
%
 This allows to rewrite Faraday's law by making the mass coefficient $e_{a}$ and the source function $f$ equal zero, whilst the damping coefficient $d_{a}$ and the conservative flux function $\Gamma$ are rewritten in the matrix form:
 %
\begin{eqnarray}\label{Eq_4}
\begin{vmatrix}
\mu_0~0\\
0~\mu_0
\end{vmatrix}
\cdot \partial_{t} |H_{x}, H_{y}|^T
+ 
\begin{vmatrix}
\partial_{x}, \partial_{y}
\end{vmatrix}
\cdot
\begin{vmatrix}
0~-E_z\\
E_z~0
\end{vmatrix}
=0 \, ,
\end{eqnarray}
%
with $H_{x}$ and $H_{y}$ the main variables of the system to be computed at the different elements of the meshed domains.

Within the 2D illustration of the cable designs shown at \autoref{Fig_1}, the position of each tape can be seen to be evenly distributed across their layers, alongside the magnetic profile derived from cables \textit{(i)} and \textit{(ii)}. Due to the experimental setup of cable \textit{(i)} incorporating an applied transport current within the shield, a blatant difference in the magnetic fields of both cables is shown at the same time instance, i.e., at $t=0.025~\mathrm{s}$, time when the transport current reach its maximum value within hysteretic (cyclic) conditions. This shows that the applied transport current causes the maximum magnetic field to be produced on the outer layer of the shield, whereas the peak magnetic field of cable \textit{(ii)} is evidently on the outermost tapes of the core.

Under both scenarios, and for a moderate transport current, i.e., $I_{tr}=0.5 I_{c}$ (per tape), being $I_{c}$ the minimum average critical current allowed for the individual 2G-HTS at the corresponding cable cores, $72.8~\mathrm{A}$ or $81~\mathrm{A}$ for the cable designs \textit{(i)} and \textit{(ii)} respectively (see tables~\ref{Tab:1} \& \ref{Tab:2}), the peak magnetic field tend towards the upper end of the spectrum with cable \textit{(i)} reaching $0.05~\mathrm{T}$ and cable \textit{(ii)} slightly higher at $0.06~\mathrm{T}$. The space between the core and the shield shows a peak of $0.015~\mathrm{T}$ when a current is applied to the shield in cable \textit{(i)}, whereas when this `shielding' current is not added for cable \textit{(ii)}, the magnetic field increases about three times, i.e, with a maximum field seen of about $0.045~\mathrm{T}$. 
 
\begin{figure}[htbp]
\centering
\resizebox{0.489\textwidth}{!}{\includegraphics{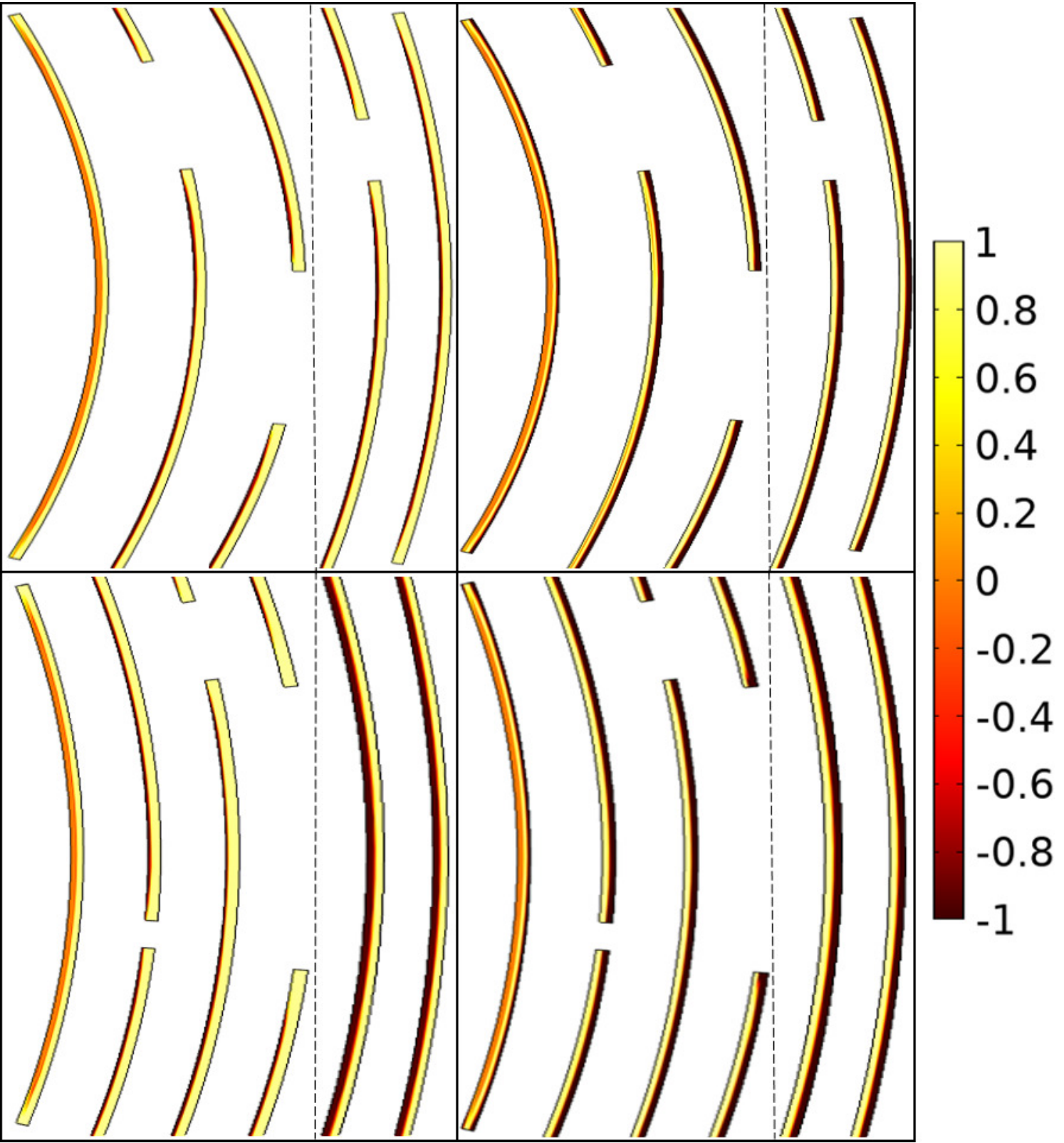}}
\caption{\label{Fig_2} {Normalized current density profiles $J_{z}/J_{c}$ for a representative section of the cables depicted in \autoref{Fig_1}. Note that the corresponding profiles for the 5-layers cable [design \textit{(i)}] correspond to the two plots at the top pane, whilst the results for the 6-layers cable [design \textit{(ii)}] are displayed underneath. The left pane corresponds to measurements at $t=0.025~\mathrm{s}$, i.e., when $I_{tr}$ is at its maximum value, showing direct evidence of transport and magnetization currents concomitantly occurring. Analogously, the right pane shows the current density profiles when $I_{tr}=0~\mathrm{A}$ at $t=0.03~\mathrm{s}$, i.e., when only hysteretic magnetization currents exist. The vertical dashed lines separates the snapshots for the `core' and `shield' layers, where the distancing between layers has been reduced for illustration.}
}
\end{figure}

To assist with the understanding of the complex electrodynamics exhibited by individual tapes at these fairly large cable designs, and by considering a sufficiently large representative section (with angular symmetry) of the corresponding two cables in \autoref{Fig_1}, the normalized current distribution profiles $J_{z}/J_{c}$ for different transport current conditions are shown in \autoref{Fig_2}. On the one hand, at $t=0.025~\mathrm{s}$, i.e., when the transport current is at its positive peak value within the first hysteresis cycle, both of the cable configurations provide similar results within the core layers, i.e., showing the concomitant action of transport and magnetization currents as $I_{tr}<I_{c}$. This shows similar features to the ones observed in HTS monofilament rounded wires~\cite{Ruiz2012APL,Ruiz2013JAP}, where the local dynamics of profiles of current density for multiple experimental conditions have been thoroughly discussed.
Thus, as it can be seen within the innermost layer (core layer 1), the +$J_{c}$ region is about but less than half the size perceptible within the other core layers. In fact, as the tapes at this layer are barely affected by the edge fields created by the neighboring tapes (due to the symmetry of the problem), a flux free region is still noticeable at the innermost region, where only the tape edges are being slightly affected by the concurrent action of magnetization currents. However, as the core layer 1 induces a magnetic field to the other layers, for instance to the core layers 2 and 3, the flux free region is not any longer observed in those, but on the contrary this is fulfilled mostly by magnetization currents flowing in the opposite direction, i.e, with -$J_{c}$. In finer detail, for the cable design \textit{(i)}, due to the transport current being applied to both, the `core' and `shield' layers, the 2G-HTS tapes at the two outermost layers of the core shown the same kind of current profiles than the shield layers, this because the dominant mechanism is the transport current. However, for the cable design \textit{(ii)}, it is to be noted than the current distribution profiles at the `shield' layers refers only to magnetization currents, reason why the distribution between positive an negatives profiles of current density is fully balanced.

On the other hand, as the transport current is reduced to $0~\mathrm{A}$, it is to be noted that the region of no flux within the core layer 1 remains the same. This evidences that we are certainly within an hysteretic behavior, i.e., beyond any magnetic relaxation events occurring mostly during the 1st half cycle of the applied current. In consequence, the remaining area of the superconducting tapes at this layer (core layer 1) are now governed by magnetization currents, i.e., showing the same amount of positive and negative negative currents along its cross section. Notice too that the amount of magnetization currents in this layer is much smaller than the one seen in all the other layers, where a 50/50 arrangement of $\pm J_{c}$ profiles can be seen. This is because any tape at the innermost layer is only affected by the edge effect (or mutual inductance) of the tapes wound in this layer,  whilst by Faraday's law the outer layers see the magnetic field created by any layer with a lower radius.


\begin{figure}[!t]
\centering
\resizebox{0.489\textwidth}{!}{\includegraphics{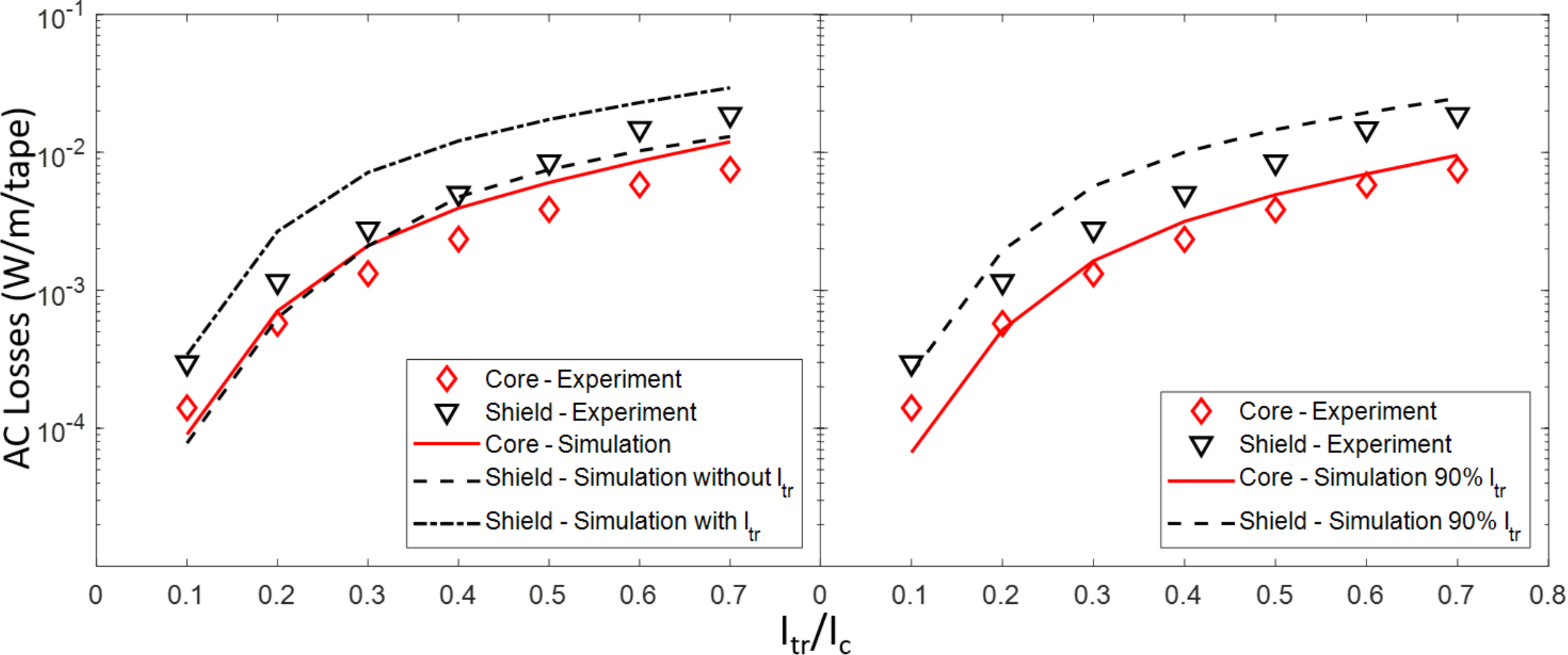}}
\caption{\label{Fig_3} {Experimental and simulated (Sim) AC losses for the cable design \textit{(i)},  with and without applied current on the `shield' layers (left pane, and with a $10\%$ reduction on the critical current measured before operational testing.}}
\end{figure}

\section{AC Losses}\label{Sec.3}

The experimental measurement of the AC-losses in large scale multilayer cables as the ones presented in~\autoref{Fig_1}, is a very demanding and complicate task. Up to know this has been achieved only by means of elaborate electrical transport measurements~\cite{Fetisov2018IEEE, Fetisov2020IOP}, which
are known to offer an adequate solution for isolated tapes or monolayer compact cables. However, for multilayer cables where the occurrence of magnetization currents and therefore magnetization losses (invisible to the electrical method), become of utter relevance, the electrical method offers only an approximate but generally lower representation of the true AC losses of the power cable. Thus, until an AC loss calorimetric rig for the testing of these cables is not completed (currently undergoing at the VNIIKP facilities), a proper estimation of the AC losses in triaxial cables can be only met by benchmarking the experimental results already available with the predictions made by numerical methods.

Thus, starting with he cable design (i), where transport current was applied to the `shield' layers, it is to be noticed that although the predicted AC-losses of the `core' layers is the same regardless whether the `shield' layers are connected in series or not to the $I_{tr}$ (see solid line at the left pane of \autoref{Fig_3}), the losses calculated at the shield layers without transport current (dashed line) are lower than the experimental ones ($\nabla$), and in fact these are nearly the same than the ones exhibited by the core layers ($\Diamond$). Nevertheless, when transport current is considered at the `shield' layers as defined in the actual experimental conditions~\cite{Fetisov2018IEEE}, we have found that for moderate current amplitudes $(0.4 I_{c} - 0.6 I_{c})$, the AC losses of the cable, including magnetization losses, can be about $20\%$ over the ones estimated by the electrical method (having as relative reference $I_{tr}$), i.e., about $10\%$ over the claimed `acceptability' window. However, this is not to be understood as a criticism to the scope of the experiments within the electrical method, nor to be inferred as a robustness of the numerical method, but simply as a more detailed benchmark with which the actual AC losses of multilayer cables can be informed. In fact, to demonstrate this and show how this $10\%$ effect can somehow alter the perception of either the experimental or numerical results, we have run a further set of simulations where the critical current is deliberately reduced in a $10\%$ (see the right pane at \autoref{Fig_3}), showing how this leads to acceptable results at both, the core and shield layers. 

\begin{figure}[!t]
\centering
\resizebox{0.489\textwidth}{!}{\includegraphics{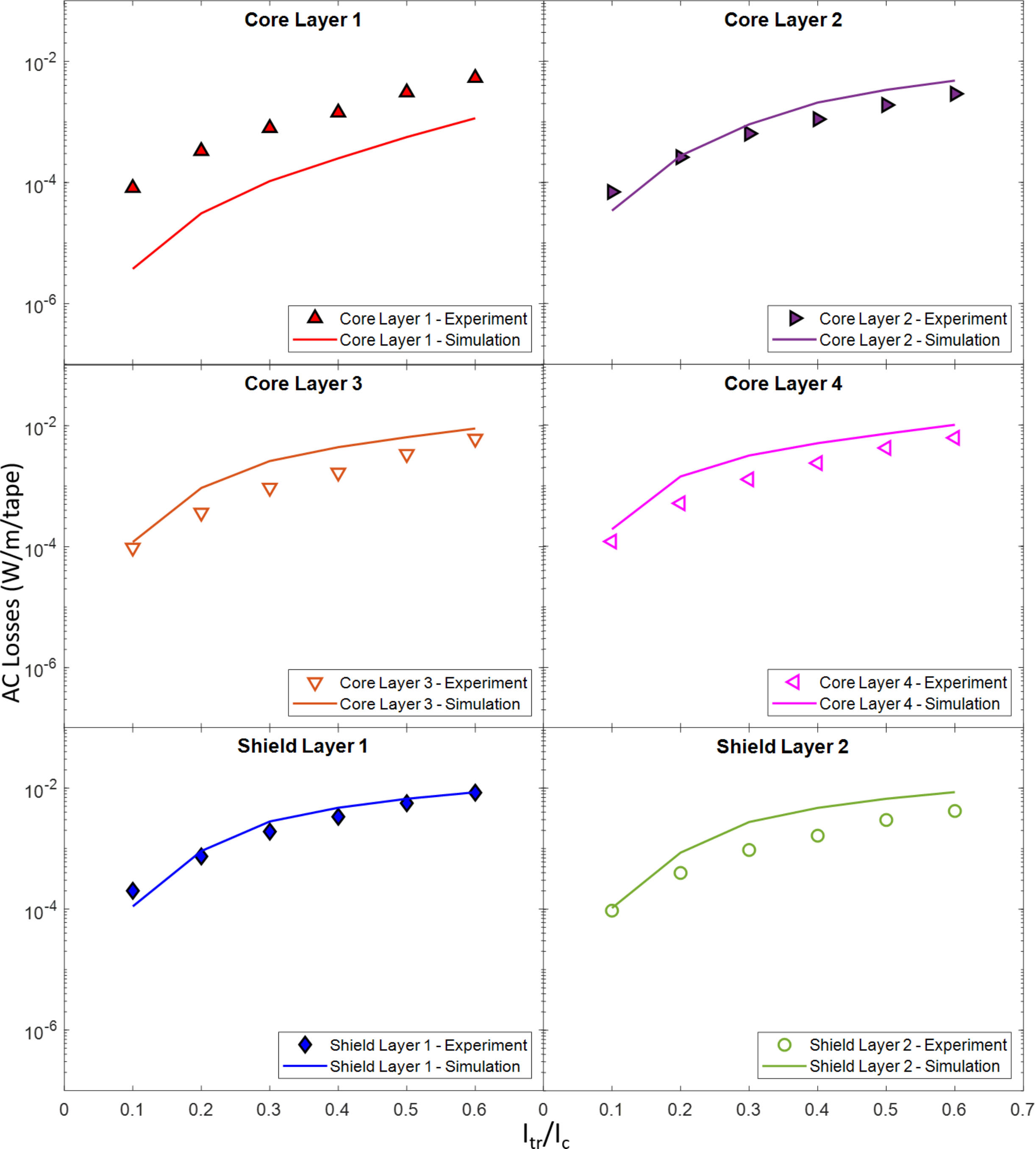}}
\caption{\label{Fig_4} {A layer-by-layer AC loss profile of the quad-layer core, bi-layer shield cable showing the results of the experimental data alongside the simulated AC losses.}}
\end{figure}

On the other hand, concerning to the cable design \textit{(ii)}, the experimental measurements of the AC losses were reported at a layer-by-layer level~\cite{Fetisov2020IOP}, what allows a better insight and assessment of the numerical model, and how it can offer a better discerning of the physical mechanisms that cause the energy losses, and influences the electromagnetic performance of the overall cable. Our results are shown in \autoref{Fig_4}, where two contrasting scenarios can be seen at first sight. On the one hand, there is a clear discrepancy between the experimental measurements obtained at the innermost layer of the cable, i.e., for the core layer 1, whilst on the other hand, their is an outstanding agreement between the experimental results and the numerically calculated AC losses for all the other five layers, including the couple of outer layers at the `shield'. Thus, although at first, several hypothesis on the numerical model conditioners were questioned to determine the origin of such increment in the AC losses for the innermost layer of the cable, in a close inspection of the experimental results, it can be seen that the measured AC losses at this layer is unexpectedly higher than the losses at the other `core' layers. However, as discussed before for the case of the cable design \textit{(i)}, the innermost layer is the one less prone to be affected by magnetization currents, i.e., with its superconducting AC losses being determined mostly by the flow of transport current, whilst for all the other layers, the AC losses is the adding result of the concomitant action between transport and magnetization currents. Therefore, from fundamental physical principles, the superconducting energy losses for the core layers 2 to 4 cannot be greater than the superconducting energy losses at the core layer 1. This results in the conclusion that the experimental measurement for the AC losses at the core layer 1 reported at \cite{Fetisov2020IOP}, is picking up a further source of energy losses that is not due to the hysteresis of the superconductor. Then, bearing in  mind that in the experiment all superconducting layers and the former were electrically isolated, as it was the case for the cable design \textit{(i)} where such behaviour is not observed, the most likely cause for this unexpected increment in the AC losses is suspected to be a defective soldering at one of the voltage taps used over these tapes. In this sense, the above does not really compromise the readability and validness of the experimental results with the electrical method for any of the layers of the cable, but on the contrary, by benchmarking with our numerical model, it offers a good and reliable testing for the electromagnetic performance of multilayer cables within well defined levels of tolerance.


\section{Conclusions}\label{Sec.4}

In this article, we have proven that by adequately setting the physical parameters of 2G-HTS tapes wound in triaxial cable configurations, and by taking advantage of two-dimensional computational frameworks in the so-called H-formulation, a sufficiently accurate estimation of the AC-losses for such complex cable arrangements can be made. Moreover, and advantage of this methodology is not only its shorter computing times, if compared with equivalent 3D models, but also the  possibility to access the local electrodynamics for the profiles of current density inside each one of the wound superconducting tapes, where mechanisms such as occurrence and consumption of magnetization currents can be seen by the dynamics of the transport currents in Bean-like profiles.

Our conclusions are supported on the study of two triaxial cable designs which have been experimentally tested at the VNIIKP facilities under different transport current conditions~\cite{Fetisov2018IEEE,Fetisov2020IOP}, with which we have established a proper benchnmark for the forecasting of the overall AC losses of 2G-HTS cold dielectric triaxial cables. Thus, alongside a critical analysis of the physical mechanisms involved in the hysteresis cycle of the superconducting tapes, by direct comparison between our numerical results with the experimental measurements achieved by the electrical testing method, we have concluded that an acceptable margin of difference for either very low transport currents $(I_{tr}\lesssim0.3I_{c})$ or very high transport currents $(I_{tr}\gtrsim0.7I_{c})$, i.e., where the impact of magnetization currents is low or overcome by the transport current losses, is of about just $10\%$. However, at moderate transport currents, which can be classified to be between the aforementioned boundaries, the impact of the magnetization currents is significant, what might lead to differences of up to $20\%$ the losses measured by the electrical testing method, as the latter is unable to give accountancy for the losses created by the magnetization currents. 


\vspace*{0.5cm}
\balance
\bibliographystyle{IEEEtran}
\bibliography{References_Ruiz_Group}

\providecommand{\noopsort}[1]{}\providecommand{\singleletter}[1]{#1}%
\begin{thebibliography}{10}
\providecommand{\url}[1]{#1}
\csname url@samestyle\endcsname
\providecommand{\newblock}{\relax}
\providecommand{\bibinfo}[2]{#2}
\providecommand{\BIBentrySTDinterwordspacing}{\spaceskip=0pt\relax}
\providecommand{\BIBentryALTinterwordstretchfactor}{4}
\providecommand{\BIBentryALTinterwordspacing}{\spaceskip=\fontdimen2\font plus
\BIBentryALTinterwordstretchfactor\fontdimen3\font minus
  \fontdimen4\font\relax}
\providecommand{\BIBforeignlanguage}[2]{{%
\expandafter\ifx\csname l@#1\endcsname\relax
\typeout{** WARNING: IEEEtran.bst: No hyphenation pattern has been}%
\typeout{** loaded for the language `#1'. Using the pattern for}%
\typeout{** the default language instead.}%
\else
\language=\csname l@#1\endcsname
\fi
#2}}
\providecommand{\BIBdecl}{\relax}
\BIBdecl

\bibitem{gilbertson2000electrical}
O.~I. Gilbertson, \emph{Electrical cables for power and signal
  transmission}.\hskip 1em plus 0.5em minus 0.4em\relax Wiley-Interscience,
  2000.

\bibitem{PamidiS.2015HsHp}
S.~Pamidi, C.~Kim, and L.~Graber, ``\BIBforeignlanguage{eng}{High-temperature
  superconducting (hts) power cables cooled by helium gas},'' in
  \emph{\BIBforeignlanguage{eng}{Superconductors in the Power Grid: Materials
  and Applications}}, 2015, pp. 225--260.

\bibitem{PrusseitWerner2015PASC}
W.~Prusseit, R.~Bach, and J.~Bock, ``\BIBforeignlanguage{eng}{Power
  applications: Superconducting cables},'' in
  \emph{\BIBforeignlanguage{eng}{Applied Superconductivity: Handbook on Devices
  and Applications}}, 2015, vol.~2, pp. 603--615.

\bibitem{Doukas2019IEEE}
D.~I. Doukas, ``\BIBforeignlanguage{English}{Superconducting transmission
  systems: Review, classification, and technology readiness assessment},''
  \emph{\BIBforeignlanguage{English}{IEEE Transactions on Applied
  Superconductivity}}, vol.~29, no.~5, pp. 1--5, Aug 2019.

\bibitem{Ruiz2018aIEEE}
B.~C. Robert and H.~S. Ruiz, ``Electromagnetic response of dc type-ii
  superconducting wires under oscillating magnetic excitations,'' \emph{IEEE
  Transactions on Applied Superconductivity}, vol.~28, no.~4, p. 8200905, 2018.

\bibitem{Ruiz2018bIEEE}
------, ``Magnetization profiles of ac type-ii superconducting wires exposed to
  dc magnetic fields,'' \emph{IEEE Transactions on Applied Superconductivity},
  vol.~28, no.~4, p. 8200805, 2018.

\bibitem{Ruiz2018SUST}
------, ``Magnetic characteristics and ac losses of dc type-ii superconductors
  under oscillating magnetic fields,'' \emph{Superconductor Science and
  Technology}, 2018.

\bibitem{Ruiz2012APL}
H.~S. Ruiz, A.~Bad\'{\i}a-Maj\'{o}s, Y.~A. Genenko, H.~Rauh, and S.~V.
  Yampolskii, ``Superconducting wire subject to synchronous oscillating
  excitations: Power dissipation, magnetic response, and low-pass filtering,''
  \emph{Applied Physics Letters}, vol. 100, no.~11, p. 112602, 2012.

\bibitem{Ruiz2013IEEE}
H.~S. Ruiz, A.~Bad{\'{\i}}a-Maj{\'{o}}s, Y.~A. Genenko, and S.~V. Yampolskii,
  ``Strong localization of the density of power losses in type-{II}
  superconducting wires,'' \emph{IEEE Transactions on Applied
  Superconductivity}, vol.~23, no.~3, pp. 8\,000\,404--8\,000\,404, June 2013.

\bibitem{Ruiz2013JAP}
H.~S. Ruiz and A.~Bad\'{\i}a-Maj\'{o}s, ``Exotic magnetic response of
  superconducting wires subject to synchronous and asynchronous oscillating
  excitations,'' \emph{Journal of Applied Physics}, vol. 113, no.~19, p.
  193906, 2013.

\bibitem{Rey2015}
A.~Malozemoff, J.~Yuan, and C.~Rey, ``5 - high-temperature superconducting
  (hts) ac cables for power grid applications,'' in \emph{Superconductors in
  the Power Grid}, ser. Woodhead Publishing Series in Energy, C.~Rey, Ed.\hskip
  1em plus 0.5em minus 0.4em\relax Woodhead Publishing, 2015, pp. 138--145.

\bibitem{Fetisov2020}
S.~S. Fetisov, V.~V. Zubko, S.~Y. Zanegin, A.~A. Nosov, and V.~S. Vysotsky,
  ``Compact 2g {HTS} power cable: new cold tests results,'' \emph{Journal of
  Physics: Conference Series}, vol. 1559, no.~1, p. 012081, jun 2020.

\bibitem{Kalsi2011SC}
S.~S. Kalsi, \emph{Applications of High Temperature Superconductors to Electric
  Power Equipment}.\hskip 1em plus 0.5em minus 0.4em\relax Hoboken, N.J:
  Wiley-IEEE Press, 2011.

\bibitem{Willen2005CIRED}
D.~Will\'en, C.~Matheus, D.~Lindsay, and M.~Gouge, ``The application of
  triaxial high-temperature superconducting power cables in distribution
  networks,'' in \emph{CIRED 2005 - 18th International Conference and
  Exhibition on Electricity Distribution}, 2005, pp. 1--4.

\bibitem{Fetisov2018IEEE}
S.~S. Fetisov, V.~V. Zubko, S.~Y. Zanegin, A.~A. Nosov, and V.~S. Vysotsky,
  ``\BIBforeignlanguage{eng}{Numerical simulation and cold test of a compact 2g
  hts power cable},'' \emph{\BIBforeignlanguage{eng}{IEEE transactions on
  applied superconductivity}}, vol.~28, no.~4, pp. 1--5, 2018.

\bibitem{Fetisov2020IOP}
S.~S. Fetisov, V.~V. Zubko, S.~Zanegin, A.~A. Nosov, and V.~S. Vysotsky,
  ``\BIBforeignlanguage{eng}{Compact 2g hts power cable: new cold tests
  results},'' vol. 1559, no.~1, 2020.

\bibitem{Fetisov2007IEEE}
V.~Sytnikov, V.~Vysotsky, A.~Rychagov, N.~Polyakova, I.~Radchenko, K.~Shutov,
  E.~Lobanov, and S.~Fetisov, ``\BIBforeignlanguage{eng}{The 5 m hts power
  cable development and test},'' \emph{\BIBforeignlanguage{eng}{IEEE
  transactions on applied superconductivity}}, vol.~17, no.~2, pp. 1684--1687,
  2007.

\bibitem{Fetisov2009IEEE}
V.~Sytnikov, V.~Vysotsky, A.~Rychagov, N.~Polyakova, I.~Radchenko, K.~Shutov,
  S.~Fetisov, A.~Nosov, and V.~Zubko, ``\BIBforeignlanguage{eng}{30 m hts power
  cable development and witness sample test},''
  \emph{\BIBforeignlanguage{eng}{IEEE transactions on applied
  superconductivity}}, vol.~19, no.~3, pp. 1702--1705, 2009.

\bibitem{Fetisov2016IEEE}
S.~S. Fetisov, V.~V. Zubko, S.~Y. Zanegin, A.~A. Nosov, V.~S. Vysotsky,
  A.~Kario, A.~Kling, W.~Goldacker, A.~Molodyk, A.~Mankevich, V.~Kalitka,
  A.~Adamenkov, S.~Samoilenkov, and D.~Melyukov,
  ``\BIBforeignlanguage{eng}{Development and characterization of a 2g hts
  roebel cable for aircraft power systems},''
  \emph{\BIBforeignlanguage{eng}{IEEE transactions on applied
  superconductivity}}, vol.~26, no.~3, pp. 1--4, 2016.

\bibitem{Fetisov2017IEEE}
S.~S. Fetisov, V.~V. Zubko, S.~Y. Zanegin, A.~A. Nosov, S.~M. Ryabov, and V.~S.
  Vysotsky, ``\BIBforeignlanguage{English}{Study of the first russian triaxial
  hts cable prototypes},'' \emph{\BIBforeignlanguage{English}{IEEE Transactions
  on Applied Superconductivity}}, vol.~27, no.~4, pp. 1--5, Jun 2017.

\bibitem{Fetisov2021IEEE}
S.~S. Fetisov, V.~S. Zubko, S.~Y. Zanegin, A.~A. Nosov, and V.~S. Vysotsky,
  ``\BIBforeignlanguage{eng}{Optimization and cold test of a triaxial 2g hts
  power cable with high current capacity},''
  \emph{\BIBforeignlanguage{eng}{IEEE transactions on applied
  superconductivity}}, vol.~31, no.~5, pp. 1--4, 2021.

\bibitem{Ruiz2019MDPI}
B.~C. Robert, M.~U. Fareed, and H.~S. Ruiz, ``How to choose the superconducting
  material law for the modelling of 2g-hts coils,'' \emph{Materials}, vol.~12,
  no.~17, p. 2679, 2019.

\bibitem{Ruiz2019JAP}
------, ``Local electromagnetic properties and hysteresis losses in uniformly
  and non-uniformly wound superconducting racetrack coils,'' \emph{Journal of
  applied physics}, vol. 126, no.~12, p. 123902, 2019.

\bibitem{Ruiz2019IOP}
------, ``Flux front dynamics and energy losses of magnetically anisotropic
  2g-{HTS} pancake coils under prospective winding deformations,''
  \emph{Engineering Research Express}, vol.~1, no.~1, p. 015037, sep 2019.

\bibitem{Ruiz2019IEEE}
M.~U. {Fareed}, B.~C. {Robert}, and H.~S. {Ruiz}, ``Electric field and energy
  losses of rounded superconducting/ferromagnetic heterostructures at
  self-field conditions,'' \emph{IEEE Transactions on Applied
  Superconductivity}, vol.~29, no.~5, p. 5900705, Aug 2019.

\bibitem{Clegg2022IOP}
M.~Clegg, M.~U. Fareed, M.~Kapolka, and H.~S. Ruiz,
  ``\BIBforeignlanguage{eng}{Computational modelling of russia\textquotesingle
  s first 2g-hts triaxial cable},'' \emph{\BIBforeignlanguage{eng}{IOP
  conference series. Materials Science and Engineering}}, vol. 1241, no.~1, p.
  012031, 2022.

\bibitem{PetrovJoP2020}
A.~Petrov, J.~Pilgrim, and I.~Golosnoy, ``\BIBforeignlanguage{eng}{2d finite
  element modelling of the ac transport power loss in multi-layer bi-2223
  cables},'' \emph{\BIBforeignlanguage{eng}{Journal of physics. Conference
  series}}, vol. 1559, no.~1, 2020.

\bibitem{SuperOx}
``Superox {2G HTS} wire. {Technical} information available at,''
  \url{http://www.superox.ru/en/products/}.

\bibitem{Zhang2018}
X.~Zhang, Z.~Zhong, J.~Geng, B.~Shen, J.~Ma, C.~Li, H.~Zhang, Q.~Dong, and
  T.~A. Coombs, ``Study of critical current and n-values of 2g hts tapes: Their
  magnetic field-angular dependence,'' \emph{Journal of Superconductivity and
  Novel Magnetism}, vol.~31, no.~12, pp. 3847--3854, 2018.

\bibitem{Fareed2022IEEE}
M.~Fareed, M.~Kapolka, B.~Robert, M.~Clegg, and H.~Ruiz,
  ``\BIBforeignlanguage{eng}{3d fem modeling of corc commercial cables with
  bean's like magnetization currents and its ac-losses behavior},''
  \emph{\BIBforeignlanguage{eng}{IEEE transactions on applied
  superconductivity}}, vol.~32, no.~4, 2022.

\end{thebibliography}
\end{document}